\documentclass[twocolumn,pre,preprintnumbers,superscriptaddress]{revtex4}
\usepackage{textcomp}
\usepackage{amsmath}
\usepackage{amssymb}
\usepackage{graphicx}
\usepackage{color}
\usepackage{soul}
\newcommand{\be}{\begin{equation}}
\newcommand{\ee}{\end{equation}}
\newcommand{\ba}{\begin{eqnarray}}
\newcommand{\ea}{\end{eqnarray}}

\newcommand{\cmp}
{\affiliation{Condensed Matter Physics Division, 
Saha Institute of Nuclear Physics, 1/AF Bidhannagar, Kolkata 700064, India.}}
\newcommand{\barasat}
{\affiliation{Barasat Government College, Barasat, Kolkata 700124, India.}}
\newcommand{\Israel}
{\affiliation{Department of Physics, Bar-Ilan University, Ramat-Gan 52900, Israel}}

\begin{document}

\title{Possible Ergodic-nonergodic regions in the quantum Sherrington-Kirkpatrick spin glass model and quantum annealing}

 \author{Sudip Mukherjee}
 \email{sudip.mukherjee@saha.ac.in}
 \barasat \cmp
 \author{Atanu Rajak}
 \email{raj.atanu009@gmail.com }
 \Israel
 \cmp
 \author{Bikas K Chakrabarti}
 \email{bikask.chakrabarti@saha.ac.in}
 \cmp

 \begin{abstract}
We explore the behavior of order parameter distribution of quantum Sherrington-Kirkpatrick 
model in the spin glass phase using Monte Carlo technique for the effective 
Suzuki-Trotter Hamiltonian at finite temperatures and that at zero 
temperature obtained using exact diagonalization method. Our numerical results indicate
the existence of low but finite temperature quantum fluctuation dominated ergodic region along with 
the classical fluctuation dominated high temperature nonergodic region in the spin glass phase of the model. 
In the ergodic region, the order parameter distribution gets narrower around the most probable 
value of the order parameter as the system size increases. In the other region, the Parisi order distribution 
function has non-vanishing value everywhere in thermodynamic limit, indicating nonergodicity. 
We also show, that the average annealing time for convergence (to a low energy level of the model; within a small error range)  
 becomes system size independent for annealing down through the (quantum fluctuation 
dominated) ergodic region. It becomes strongly system size dependent for annealing through the nonergodic region. 
Possible finite size scaling type behavior for the extent of the ergodic region is also addressed.
\end{abstract}
\maketitle

\section{Introduction}
Considerable amount of investigations have been made in studying the nonergodic
behavior \cite{sudip-binder} of the spin glass phase of the classical Sherrington-Kirkpatrick (SK) spin glass model \cite{sudip-sk}. 
The phenomenon of replica symmetry breaking, induced by nonergodicity, 
occurs due to the appearance of macroscopically high free-energy 
barriers separating the local minima. Such highly rugged nature of free-energy landscape 
in spin glass phase causes the system to get trapped into any one (locally) self-similar region of the configuration space. 
Consequently one gets a broad order parameter distribution (or the replica symmetry breaking) in the spin glass 
phase as suggested by Parisi~\cite{sudip-parisi}. 
In this case, along with the peak at any non-zero value of the order parameter, its distribution 
also contains a long tail extended to the zero value of the order parameter in the thermodynamic limit. 
This localization due to nonergodicity has been identified to be responsible for the NP hardness of equivalent 
optimization problems (see e.g., \cite{sudip-das}).

The situation seems to be quite different when the SK spin glass is placed under 
a transverse field. Due to the presence of the quantum fluctuations, the system 
is able to tunnel through the tall (but narrow) 
free-energy barriers~\cite{sudip-ray,sudip-bikas,sudip-ttc-book,sudip-lidar,sudip-troyer,sudip-Katzgraber}, 
inducing ergodicity (or absence of replica symmetry breaking). 
Consequently one would expect a narrowly peaked order parameter 
distribution in quantum SK spin glass model in the thermodynamic limit \cite{sudip-ray}. 
This ergodicity has been identified to be 
responsible (see e.g., \cite{sudip-bikas,sudip-ttc-book}) for the success of quantum annealing.

We have studied the nature of order parameter distribution of transverse field SK spin glass 
at finite temperature using Monte Carlo simulation of the effective Suzuki-Trotter Hamiltonian 
and using the exact diagonalization technique at zero temperature. In this numerical study we tried to 
identify the possible ergodic spin glass phase (due to quantum tunneling) of the system. 
We find a low temperature region in the quantum SK system, where the tails of the order 
parameter distribution vanishes in thermodynamic limit, suggesting convergence of the 
order parameter distribution to be a peaked one around  
the most probable value. Although the system sizes we studied are not 
very large, we believe our study clearly indicates the existence of a low temperature 
 ergodic region in the spin glass phase of this quantum SK model.
On the other hand, in other (high temperature) part of the spin glass phase, the order parameter 
distribution appears to remain Parisi type~\cite{sudip-young} which 
indicates lack of ergodicity in this part of the spin glass phase. We have already identified 
 \cite{sudip-sudip} the quantum fluctuation dominated part of the spin glass phase boundary of this 
 model, crossing over at finite temperature to the classical fluctuation dominated part (see also \cite{sudip-yao}).
 Here we find that the line separating the ergodic and the nonergodic regions pass through the zero 
 temperature-zero transverse field point and the above mentioned quantum-classical crossover point on 
 the phase boundary.

We also study the variation of the average annealing time in the finite temperature 
Suzuki-Trotter Hamiltonian dynamics for the model in both the ergodic and nonergodic 
regions. For annealing down to a fixed low temperature and low transverse field point
through the (quantum fluctuation dominated) ergodic region,  we find the average annealing 
time to be independent of system size. On the other hand the average annealing time 
is observed to grow strongly with the system size, when similar annealing is performed 
through the (classical fluctuation dominated) nonergodic region.

\section{Model}
\label{model}

The Hamiltonian of the quantum SK spin glass model with $N$ Ising spins is given by (see e.g.,~\cite{sudip-ttc-book})
\begin{align}
H  = H_0 + H_I;~ H_0=-\sum_{i < j} J_{ij}\sigma_i^z\sigma_j^z;~ H_I=-{\Gamma} \sum_{i = 1}^N\sigma_i^x . \label{Ham}  
\end{align}
Here $\sigma_i^z$, $\sigma_i^x$ are the $z$ and $x$ components of Pauli spin matrices respectively and $\Gamma$ denotes
the transverse field. The Hamiltonian in Eq.~(\ref{Ham}) becomes the classical SK spin glass Hamiltonian ($H_0$) for 
zero value of the transverse field.  The spin-spin couplings ($J_{ij}$) are distributed  following Gaussian distribution
$\rho (J_{ij}) = \Big (\frac{N}{2{\pi}J^2}\Big)^{\frac{1}{2}}\exp\Big (\frac{-NJ_{ij}^2}{2J}\Big)$, where  the mean and
standard deviation of the distribution are zero and $J/ \sqrt{N}$ respectively (see e.g.,~\cite{sudip-ttc-book}). In this work we
take $J = 1$. To perform Monte Carlo simulation at finite temperature we map the Hamiltonian~(\ref{Ham}) into an effective
classical Hamiltonian $H_{eff}$ by using Suzuki-Trotter formalism~(see e.g., \cite{sudip-bkc-book}):
\begin{eqnarray}
 H_{eff}& =& -\sum_{m=1}^M \sum_{i < j}^{N} \frac{J_{ij}}{M}\sigma_i^m\sigma_j^m \nonumber\\
 &-& \sum_{i=1}^N\sum_{m=1}^M\frac{1}{2\beta}\text{log~coth}\frac{\beta\Gamma}{M}\sigma_i^m\sigma_i^{m+1} \label{H_cl},
\end{eqnarray}
where $\sigma_i^m$ ($=\pm 1$) represents the $i$-th (classical) Ising spin in the $m$-th replica. We have an additional 
dimension (in Eq.~\ref{H_cl}), namely the Trotter dimension. Here $M$ denotes the total number of Trotter slices and
$\beta$ is the inverse of temperature $T$. ${M\to \infty}$ as $T\to 0$. 
\begin{figure*}[htb]
\begin{center}
\includegraphics[width=7.0cm]{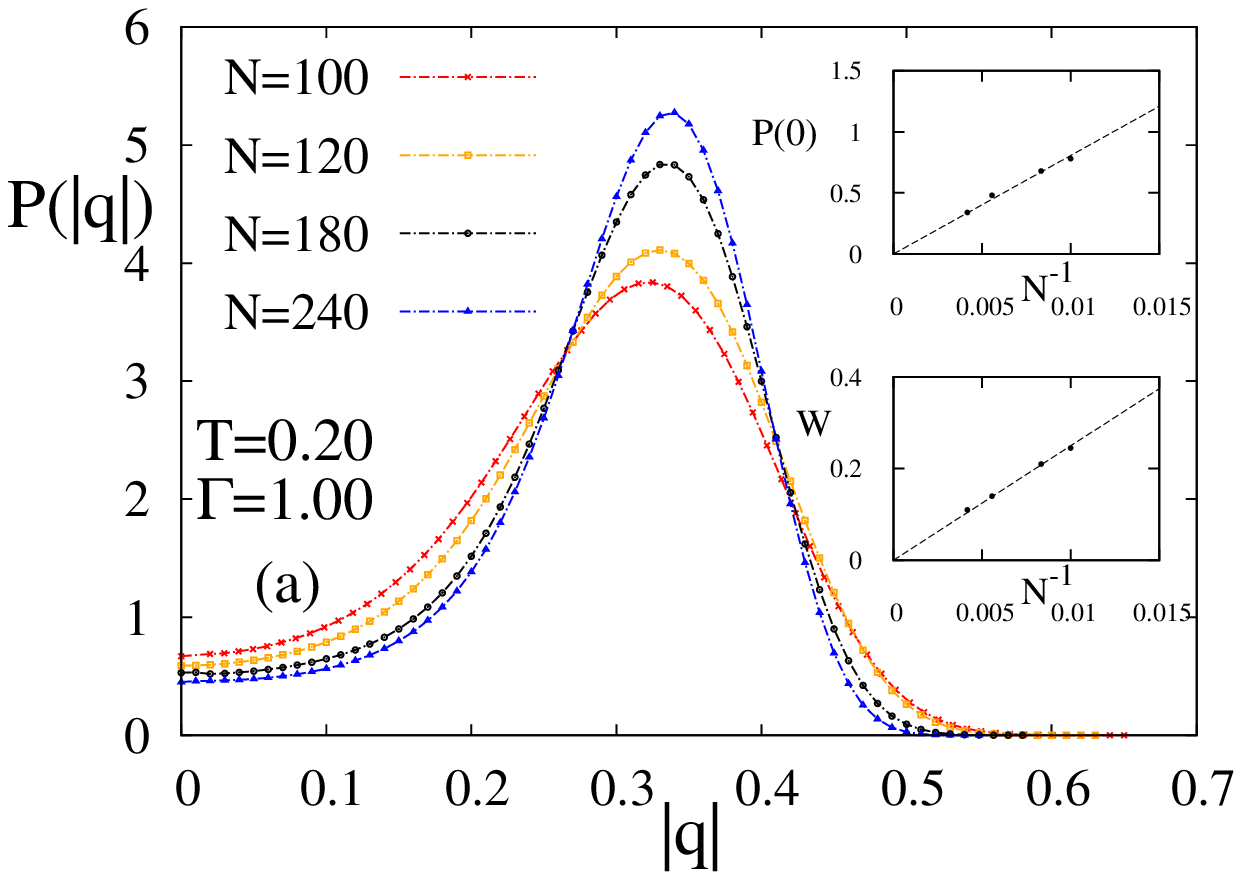}
\includegraphics[width=7.0cm]{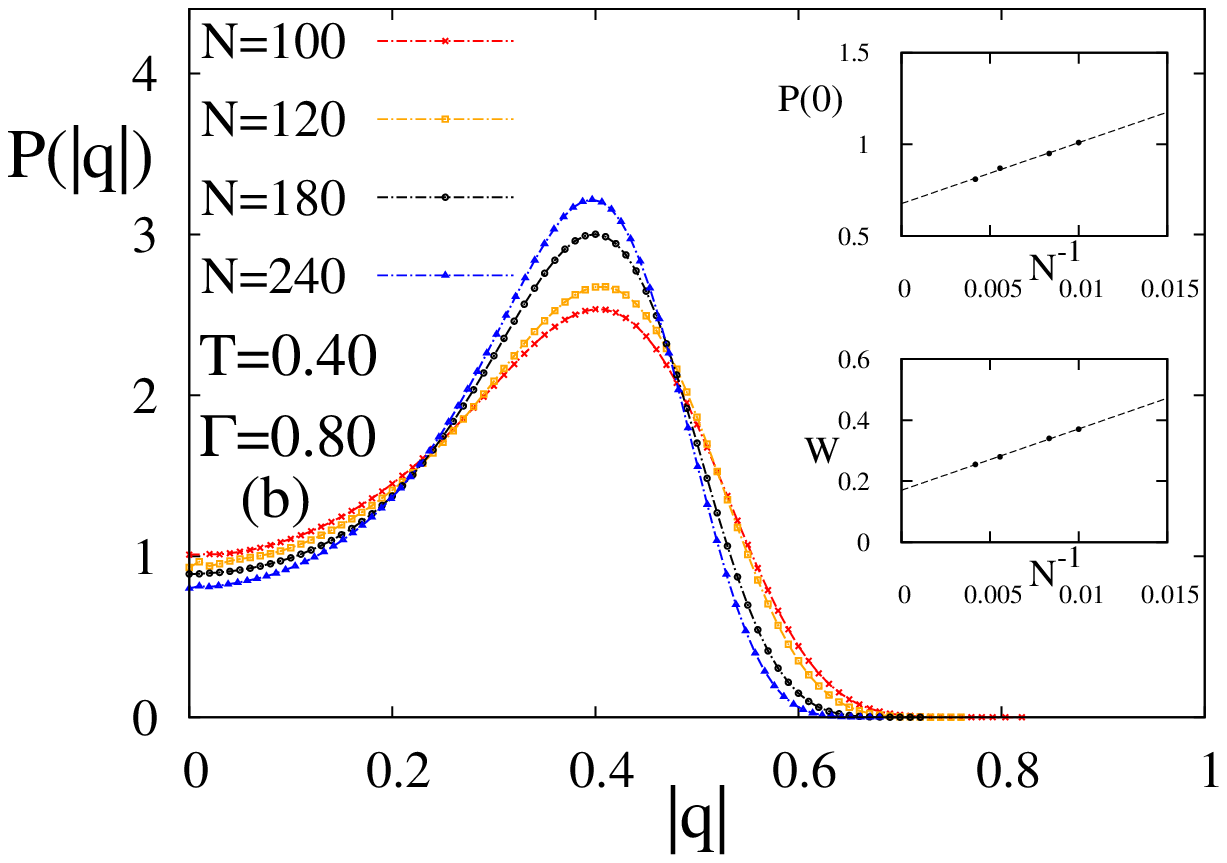}
\end{center}
\caption{(Color online) Monte Carlo results for the plots of the area-normalized order parameter distribution $P(|q|)$
for given sets of transverse field $\Gamma$ and temperature $T$ are shown: (a) for $T=0.20$ and 
$\Gamma = 1.00$, (b) for $T = 0.40$ and $\Gamma = 0.80$. Extrapolations of $P(0)$ with $1/N$ are shown in
the insets. In the first case the extrapolated value of $P(0)$ and $W$ tend to zero in the large system size limit 
whereas in the other case the values of such quantities remain finite even in thermodynamic limit.}
\label{area_ergodic}
\end{figure*}
\begin{figure*}[htb]
\begin{center}
\includegraphics[width=7.0cm]{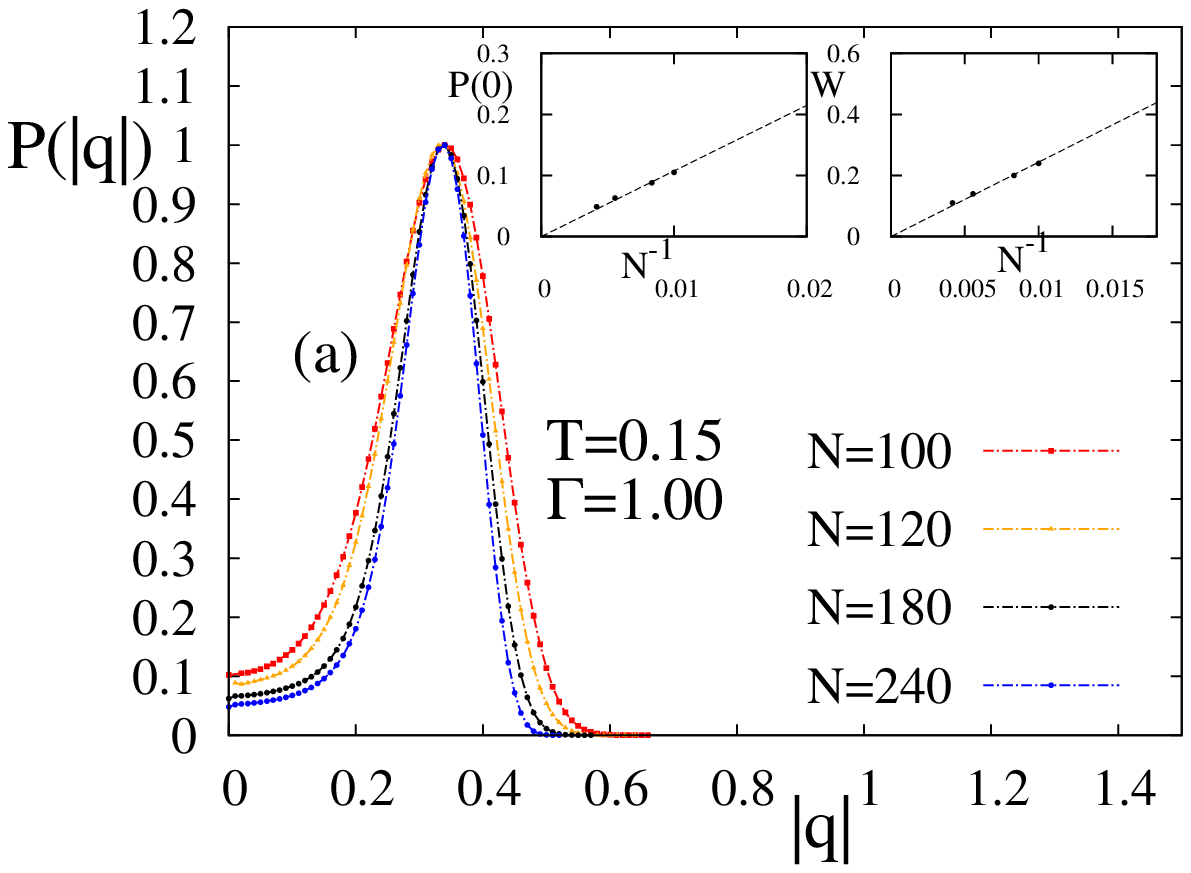}
\includegraphics[width=7.0cm]{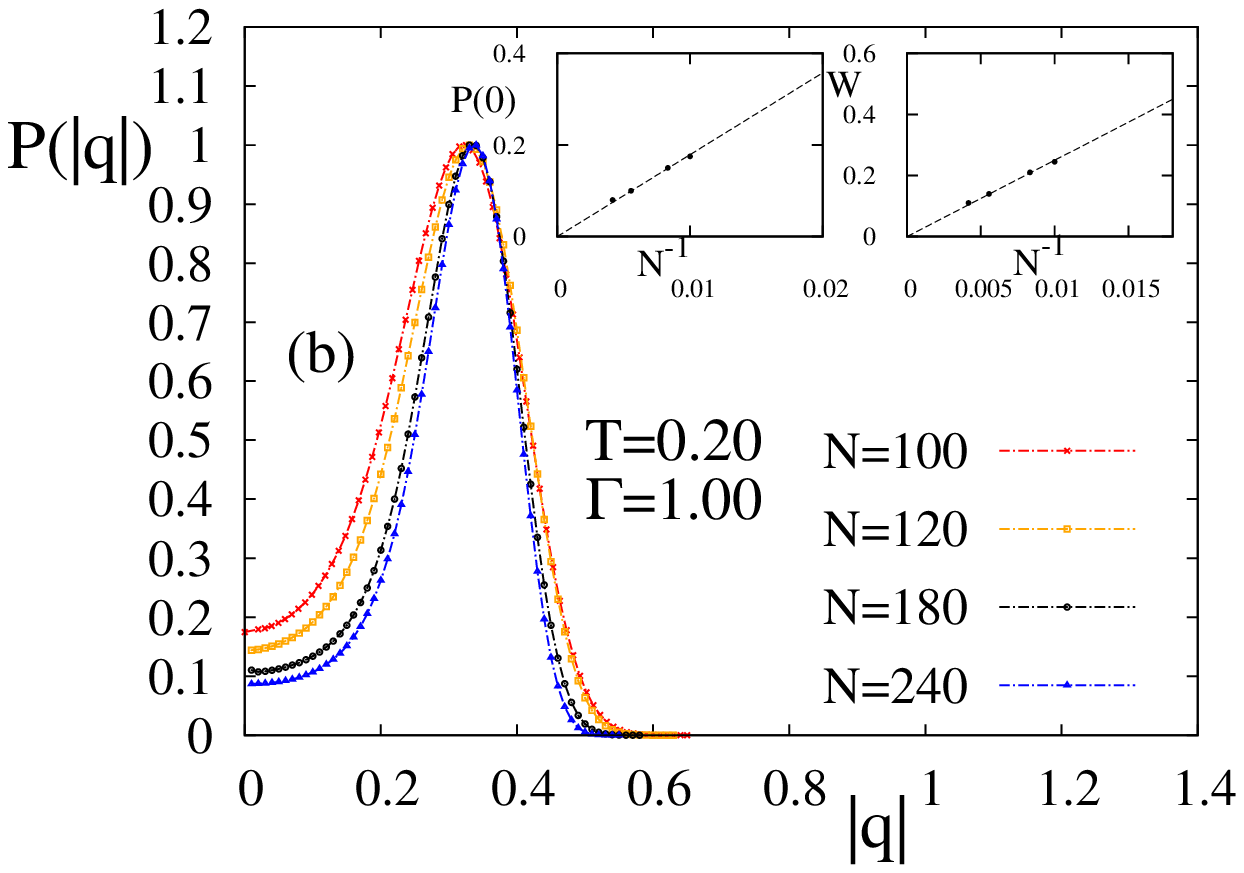}
\end{center}
\caption{(Color online) Monte Carlo results for the plots of the peak-normalized order parameter distribution $P(|q|)$ 
for given sets of transverse field $\Gamma$ and temperature $T$ are shown: (a) for $T=0.15$ and $\Gamma = 1.00$, (b) 
for $T = 0.20$ and $\Gamma = 1.00$. Extrapolations of $P(0)$ with $1/N$ are shown in the insets.
In both the cases the extrapolated value of $P(0)$ tends to zero in the large system size limit.}
\label{ergodic}
\end{figure*}
\begin{figure*}[htb]
\begin{center}
\includegraphics[width=7.0cm]{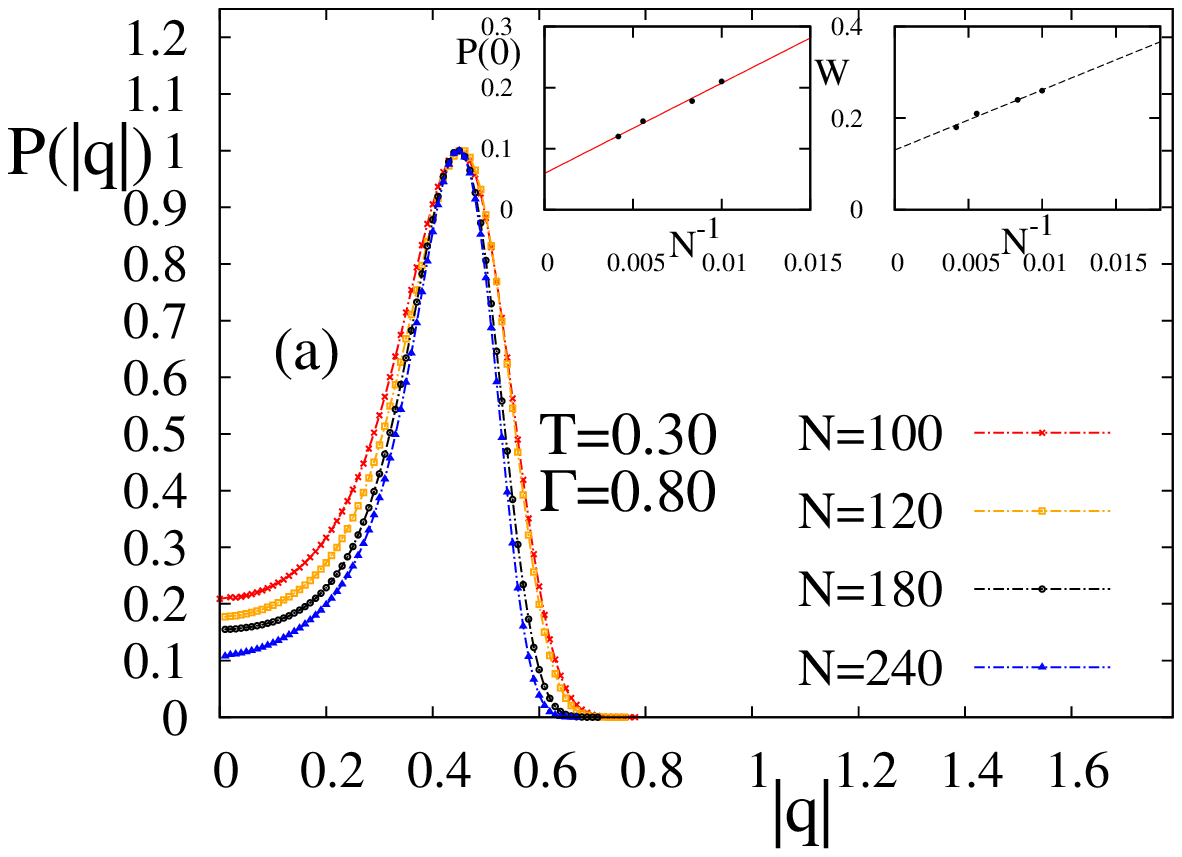}
\includegraphics[width=7.0cm]{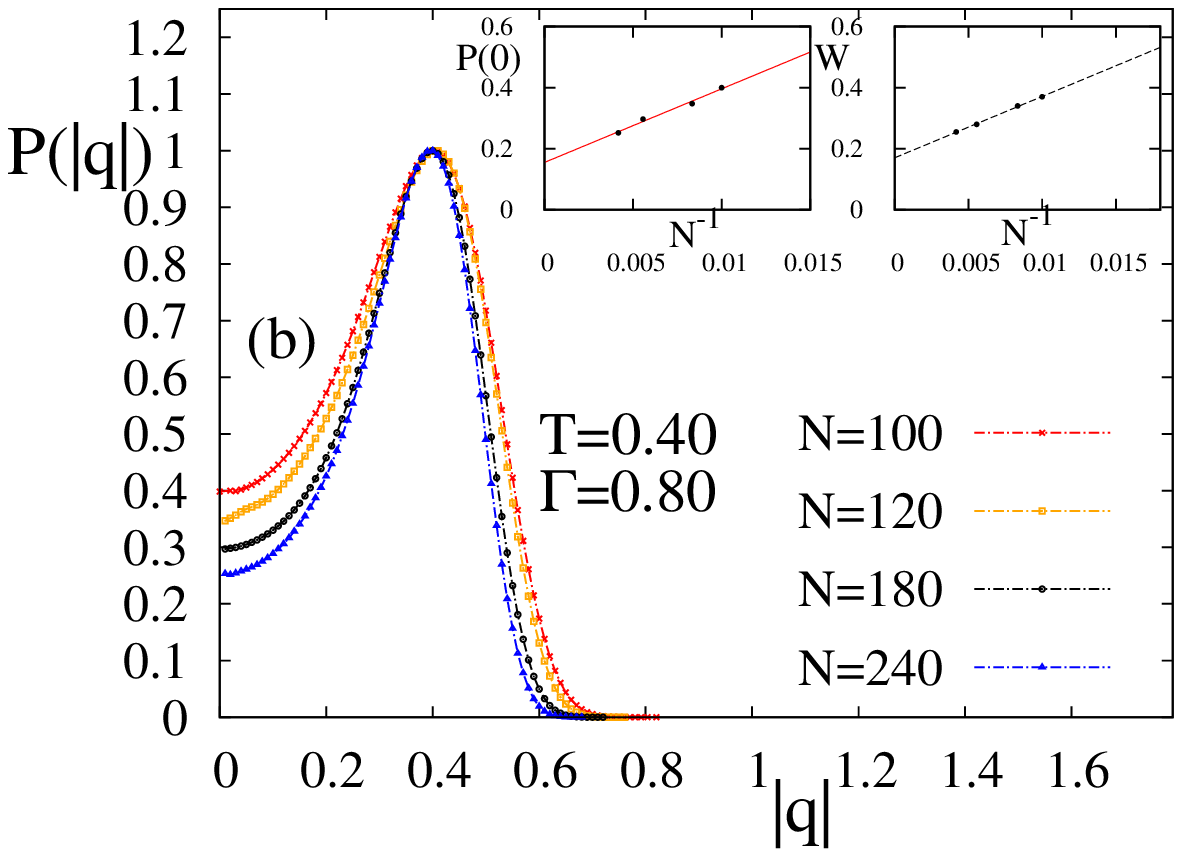}
\end{center}
\caption{(Color online) Monte Carlo results for the plots of the peak-normalized order parameter distribution $P(|q|)$ 
for given sets of transverse field $\Gamma$ and temperature $T$ are shown: (a) for $T=0.30$ and $\Gamma = 0.80$, (b) 
for $T = 0.40$ and $\Gamma = 0.80$. Again we extrapolate the values of $P(0)$ with $1/N$ which are shown in the insets. 
In these cases extrapolated values of $P(0)$ for large system size limit remain finite.}
\label{non-ergodic}
\end{figure*}

\section{Monte Carlo Results}
For finite temperature study, we perform Monte Carlo simulation on $H_{eff}$ to obtain the order parameter
distribution in spin glass phase of our model. To obtain such distribution function we first allow the system
to equilibrate with $t_0$ Monte Carlo steps and the thermal averaging is made over next $t_1$ time steps. 
In one Monte Carlo step we update all the spins of the system once. After the equilibration in each Monte Carlo
step $t$ we calculate the replica overlap $q^{\alpha \beta}(t)$, which is defined as 
$q^{\alpha \beta}(t)=\frac{1}{NM}\sum_{i=1}^N\sum_{m=1}^M(\sigma_i^m(t))^{\alpha}(\sigma_i^m(t))^{\beta}$. Here $(\sigma_i^m)^{\alpha}$ and
$(\sigma_i^m)^{\beta}$ denote the spins of two replicas (in the $m$-th Trotter slice) having identical set of
$J_{ij}$'s. The order parameter distribution $P(q)$ can be obtained as
\begin{align*}
 P(q)=\overline{ \frac{1}{t_1}\sum_{t=t_0}^{t_0+t_1}\delta(q-q^{\alpha \beta}(t)) } ,
\end{align*}
where the overhead bar denotes the configuration average over several sets of $J_{ij}$'s. The order parameter
$q$ is defined as $q = \frac{1}{MN}\sum_{m=1}^{M}\sum_{i=1}^{N}\overline{{\langle \sigma_i^m \rangle}^2}$, 
where $\langle .. \rangle$ denotes the thermal average for a given configuration of disorder. From numerical 
data we compute the distribution function $P(q)$ for a given set of $T$ and $\Gamma$ by considering 
both area normalization and peak normalization (where peaks of the distributions are normalized). 

In our simulation we work with the system sizes $N = 60, 120, 180, 240$ and the number of Trotter slices is $M=15$. 
We have found that the equilibrium time  of the system is not identical throughout the entire region of $\Gamma - T$ plane. 
The equilibrium time of the system (for $60 \leq N \leq 240$) is typically $\lesssim 10^6$ within the region $T< 0.25$ and 
$\Gamma < 0.40$, whereas it becomes $\lesssim 10^5$  for the rest of the spin glass phase region. We take 
$t_1=1.5 \times 10^{5}$ for Monte Carlo averaging and the configuration average is made over $1000$ samples (configurations). 
Because of its symmetry we have determined the distribution of $|q|$ instead of $q$. We observe that the value of $P(|q|)$ 
for $q=0$ has a clear system size dependence. We extrapolate the values of $P(0)$ with $1/N$ to get the value of 
$P(0)$ for infinite system size. We also calculate the width $W$ at half maximum of the distribution function.   
The width $W$ is define as  $W = |q_2 - q_1|$ where the value of $P(|q|)$ becomes half of its maximum value at $q= q_1$, $q_2$. 
Again we extrapolate the values of $W$ with $1/N$. We observe two distinct behaviors of such extrapolated values of both 
$P(0)$ and $W$ in spin glass phase. In the low temperature (and high transverse field) case, we notice that the 
values of $P(0)$ and $W$ both go to zero in the large system size limit (see Fig.~\ref{area_ergodic}a). Such an observation 
indicates $P(|q|)$ would approaches to Gaussian form in thermodynamic limit, suggesting ergodic behavior of the system. 
In contrast, for the other case (high temperature case) we find that $P(0)$ has finite value even in thermodynamic 
limit (see Fig.~\ref{area_ergodic}b)). There seems to be no possibility of $P(|q|)$ to approach the Gaussian form 
of distribution for infinite system size limit. It indicates that the system remains nonergodic in this region of 
spin glass phase. To identify the ergodic and nonergodic regions in spin glass phase more accurately, we also study 
the behavior of the peak normalized order parameter distribution. From such study again we find in low temperature 
and high transverse field the values of $P(0)$ and $W$ (extrapolated with $1/N$) become zero in thermodynamic limit 
(see Fig.~\ref{ergodic} (a, b)). Again from peak normalized order parameter distribution we find that for 
high temperature and low transverse field the extrapolated values of the tail and width of distribution remains
non zero in infinite system size limit (see Fig.~\ref{non-ergodic} (a, b)).

\section{Zero-temperature diagonalization results}
\label{zero_temp}
For zero temperature study of our model, we have investigated 
the distribution of the spin glass order parameter using an exact diagonalization 
technique. The diagonalization of the quantum spin glass has been performed using Lanczos algorithm~\cite{lanczos_52}
to obtain its ground state. In this case we have considered system sizes ($N$) upto $20$. 
The Hamiltonian in Eq.~(\ref{Ham}) can be written in the spin basis states which are indeed the eigenstates 
of the Hamiltonian $H_0$. After performing diagonalization, the $n$-th eigenstate of the Hamiltonian in 
Eq.~(\ref{Ham}) is found out as $|\psi_n\rangle~= \sum_{\alpha=0}^{2^{N-1}} a_{\alpha}^n |\varphi_\alpha\rangle$, 
where $a_{\alpha}^n=\langle\varphi_{\alpha}|\psi_n\rangle$ and $|\varphi_\alpha\rangle$ denote the eigenstates 
of the Hamiltonian $H_0$. As the consequence of our interest in zero temperature analysis, 
we shall here mainly focus on the ground state ($|\psi_0\rangle$) averaging of different quantities of interest. 
One can define the order parameter for this zero temperature system as 
$Q = (1/N) \sum_i \overline{\langle\psi_0|\sigma_i^z|\psi_0\rangle^2}=(1/N)\sum_i\overline{Q_i}$ (note that $Q$ here for $T = 0$ 
differs from $q$ defined earlier for $T \ne 0$, using replica average)~\cite{sudip-sudip}. 
Here also, the overhead bar indicates the configuration averaging. $Q_i$ denotes the site-dependent 
local order parameter value. The distribution of the local order parameter is then represented by 
\begin{equation}
P(|Q|)=\overline{\frac{1}{N}\sum_{i=1}^N\delta(|Q|-Q_i)}.
\end{equation}

\begin{figure*}[ht]
\begin{center}
\includegraphics[width=7.0cm]{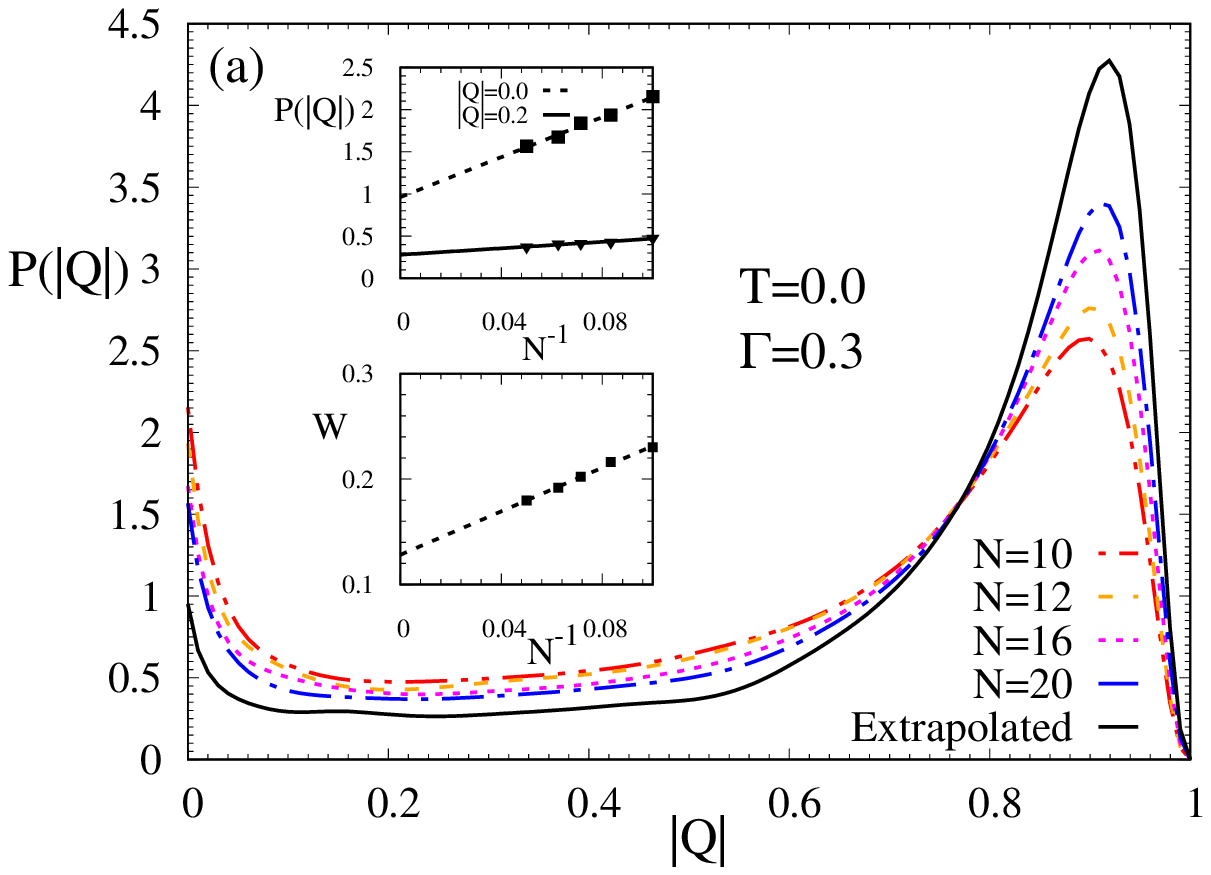}
\includegraphics[width=7.0cm]{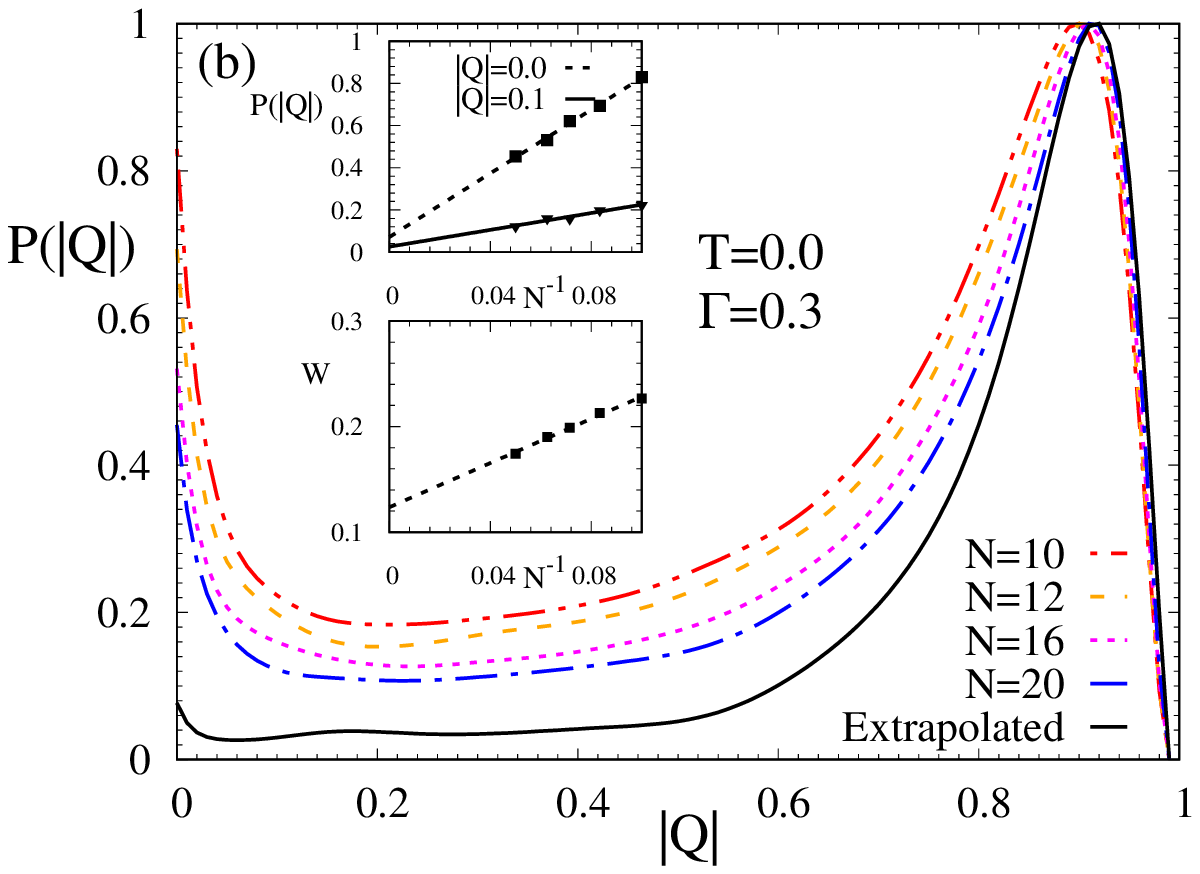}
\end{center}
\caption{(Color online) Exact diagonalization results (at zero temperature) for the 
variation of $P(|Q|)$ as a function of $|Q|$ for quantum SK spin glass 
for four different system sizes $N$ at $T=0$ and $\Gamma=0.3$ are shown. 
For (a) the area under the $P(|Q|)$ curves for each $N$ is normalized to $1$, whereas
for (b) the peaks of all $P(|Q|)$ curves are normalized to their maximum 
values. For both the plots, the extrapolated values of $P(|Q|)$
for large $N$ (see inset) are plotted against $|Q|$ as well. The typical extrapolations 
of $P(|Q|)$, for $|Q|=0.0$ and $0.1$, and $W$ with $N$ are shown in the inset of both the plots. 
 }
\label{op_dist_0.3gama}
\end{figure*}

Similar to the case of finite temperature, we here also have investigated the behavior 
of $P(|Q|)$ in the spin glass phase at different values of $\Gamma$. The variation of 
$P(|Q|)$ as a function $|Q|$ at $\Gamma=0.3$ is shown in Fig.~\ref{op_dist_0.3gama} 
for four different system sizes. It may be noted that in this case also we have plotted 
the distribution curves for different system sizes normalized to their maximum values 
as well as area normalization under the curves. 
From both the plots in Fig.~\ref{op_dist_0.3gama}, we observe that $P(|Q|)$ shows a peak at a finite value 
of $|Q|$ along with non-zero weight at $Q=0$. However, the value of $P(0)$ decreases with the
increase of the system size (although one can still detect an upward rise of $P(|Q|)$ for lower values of $|Q|$). 
To get the behavior of $P(|Q|)$ in the thermodynamic limit, we have computed the value 
of $P(|Q|)$ for infinite size system for each $|Q|$ by plotting $P(|Q|)$ as a function of $1/N$.
The extrapolation of $P(|Q|)$ for infinite size system is shown in the inset of 
both plots in Fig.~\ref{op_dist_0.3gama} for $Q=0.0$ and $0.1$. In addition, we have also
calculated the width ($W$) at half of maximum: $W=|Q_2-Q_1|$ where at $Q_2$ 
and $Q_1$ the value of $P(|Q|)$ is the half of its maximum value. We plot $W$ 
as a function of $1/N$ to get its extrapolated value for infinite size system (see Fig.~\ref{op_dist_0.3gama}).
Finally we have also plotted $P(|Q|)$ as a function of $|Q|$ with the 
extrapolated value of $P(|Q|)$ for infinite system size (see Fig.~\ref{op_dist_0.3gama}).
One can observe that $P(|Q|)$ curve for infinite system becomes narrower as compared 
to the cases of finite system size. On the other hand, due to the limitation of the maximum system size we could consider
in our numerics, we are here not able to get $P(|Q|)$ curve for very large system sizes showing results consistent with  
delta function form. The effect of the limitation 
of the system size is also present in the plot of $W$ with $1/N$ since the extrapolated $W$ does not 
acquire strictly zero value here. 
However, we infer from our extrapolated numerical analysis  
(from results of the small system sizes) that eventually the $P(|Q|)$ curve would become a delta function at a  
finite values of $|Q|$ in thermodynamic limit. This would suggest the system to become ergodic 
in the spin glass phase at zero temperature with a 
definite spin glass order parameter value.

\begin{figure}[h]
\begin{center}
\includegraphics[width=7.2cm]{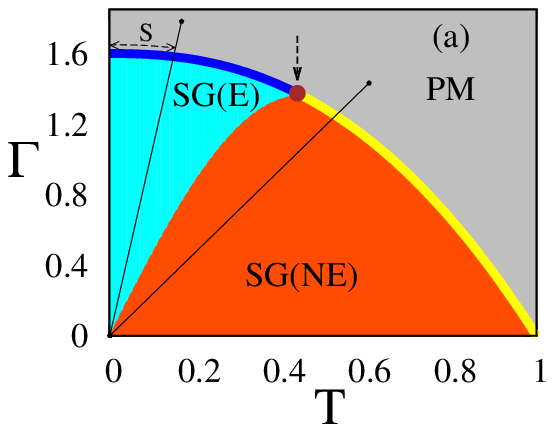}
\includegraphics[width=7.2cm]{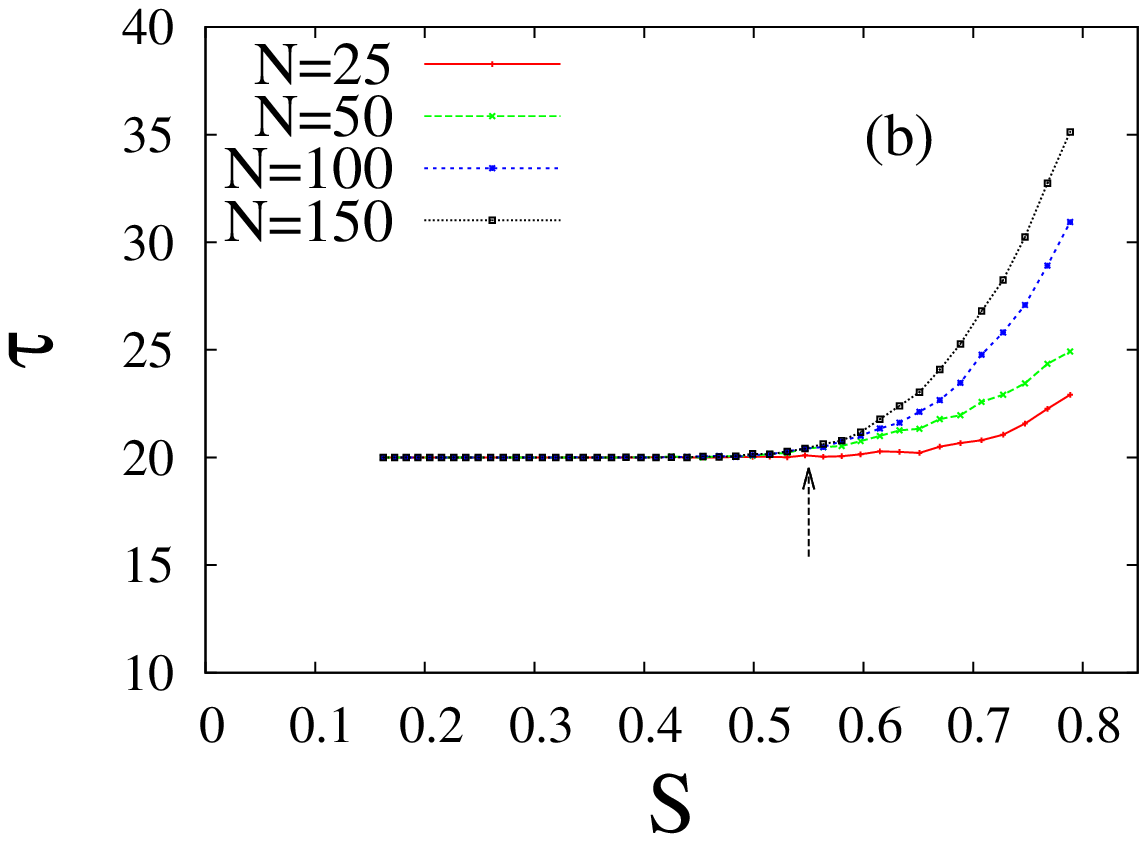}
\end{center}
\caption{(Color online) (a) Schematic phase diagram of the quantum SK model is shown (cf.~\cite{sudip-sudip}). 
Here SG and PM denote the spin glass and paramagnetic phases respectively. 
Our numerical simulations indicate that the spin glass phase is further divided 
into two regions: Ergodic SG(E) region and nonergodic SG(NE) region. 
The dot on the SG-PM phase boundary line 
indicates the quantum-classical crossover point in the critical behavior of the model~\cite{sudip-sudip,sudip-yao}. 
We anneal by tuning both $T$ and $\Gamma$ following different
linear paths passing through both SG(E) and SG(NE) regions 
(e.g., as indicated by the two inclined straight lines in the figure).
 (b) Shows the variation of annealing time $\tau$ with $S$,   
 the length of the arc along the phase boundary starting from the pure
 quantum critical point ($T=0, \Gamma \simeq 1.6$), upto the 
 crossing point of the annealing line on the phase boundary.
 One does not get any system size dependence of $\tau$ upto $S \simeq 0.55$ (corresponding to $T \simeq 0.46, \Gamma \simeq 1.35$; 
 indicated in both the figures by vertical arrows). 
 As the annealing line passes through the SG(NE) region (beyond the quantum-classical crossover point) 
 $\tau$ is seen to acquire a strong system size dependence.
 }
\label{phase_diagram}
\end{figure}

 \section{Annealing through ergodic and nonergodic regions}\label{QA_EE_NEE}
Our observations described in the earlier sections indicate clearly the existence of 
both ergodic (low temperature and high transverse field) region as well as nonergodic 
(high temperature and low transverse field) regions, separated by a line originating 
from $T =0, \Gamma = 0$ and the passing through the quantum-classical crossover point obtained earlier~\cite{sudip-sudip,sudip-yao} 
on the phase boundary of the model. In order to check the dynamical features 
of these two regions, we have studied the annealing behavior of the system, again 
using the Suzuki-Trotter effective Hamiltonian with time ($t$) dependent $T$ and $\Gamma$: 
$T(t) = T_0(1 - \frac{t}{\tau})$ and ${\Gamma}(t) = {\Gamma}_0(1 - \frac{t}{\tau})$. 
Here $T_0$ and ${\Gamma}_0$ correspond to points in the para phase such that they are 
practically equidistant from the phase boundary line in different parts of the phase diagram. 
We look for the variation  of the annealing time $\tau$ to reach a very low free-energy 
(corresponding to very small values of $T \simeq 10^{-3} \simeq \Gamma$ to avoid singularities in the effective interaction $H_{eff}$  
and dynamics; putting by hand these values of $T$ and $\Gamma$ in the last $t = \tau$ step of the annealing schedule), 
starting from para phase (high $T$ and $\Gamma$ values). We study annealing of the system when the annealing 
path (schedule) passes either through ergodic or through nonergodic region (see Fig.~\ref{phase_diagram}a). 
We find that the annealing time $\tau$ remains fairly constant for any annealing path (schedule) passing entirely 
through the ergodic SG(E) region and becomes strongly system size ($N$) dependent as the path passes through the 
nonergodic region (see Fig.~\ref{phase_diagram}b). It may be mentioned, deep inside the classical region 
of the spin glass phase (for $S$ value $\gtrsim 1$; see Fig.~\ref{phase_diagram}) the annealing time
becomes strongly configuration dependent and hence the $N$-dependence of the average value of $\tau$ becomes somewhat irregular.

\section{Summary and Discussions}
We have studied the  order parameter distribution
in the spin glass phase of the
quantum SK model, both at finite temperature (using Monte Carlo simulation of 
the effective Hamiltonian (\ref{H_cl})) and at zero temperature 
(using exact diagonalization). For Monte Carlo simulation we have taken system sizes $N=60, 120, 180, 240$ along with 
the Trotter size $M = 15$ (Figs.~\ref{ergodic},\ref{non-ergodic}). It may be mentioned that we cheeked that the Monte Carlo results  
remain practically unchanged if for such system sizes we vary the number of Trotter slices $M$ 
 with the system size $N$ keeping the value of the scaled variable $M/N^{z/d}$ constant, where $z$ denotes 
the dynamical exponent and $d$ is the effective dimension of the system (see~\cite{sudip-sudip} for details).
For zero temperature analysis we considered the system sizes 
$N=10, 12, 16$ and $20$ (Fig.~\ref{op_dist_0.3gama}). In the ergodic region SG(E) (see Fig.~\ref{phase_diagram}a), 
the (extrapolated) order parameter distribution 
is found to converge to a Gaussian form around a most probable value
with increase of the system size (see Figs.~\ref{ergodic}$(a,b)$ for $T \ne 0$). 
Although at $T = 0$ the system sizes we considered are very small, it can be anticipated that we will get 
a single and narrow peak in the order parameter distribution (see Fig.~\ref{op_dist_0.3gama}) around a most probable 
value for thermodynamically large system indicating the ergodicity of the spin glass 
phase at zero temperature.  
On the other hand, in the nonergodic region SG(NE) (see Fig.~\ref{phase_diagram}a), 
we get Parisi-type order parameter distribution where 
the long tail extends upto zero value of order parameter (see Figs.~\ref{non-ergodic}$(a,b)$). 
This (nonzero) weight of the distribution near the origin 
remains non-vanishing with increase in the system size $N$.
This behavior of the order parameter distribution indicates the absence of ergodicity in the 
system in the SG(NE) region.

These results indicate the different regions of the spin glass phase of the 
quantum SK model as shown in Fig.~\ref{phase_diagram}a. It may be noted that the line 
separating the low temperature (quantum fluctuation driven) ergodic region of the quantum spin glass phase 
from the high temperature nonergodic region passes through the quantum-classical 
crossover point on the spin glass phase boundary obtained earlier~\cite{sudip-sudip,sudip-yao}. Apart from
this low temperature part of the spin glass phase the entire para phase of course remains ergodic. 

In order to test the role of this quantum fluctuation induced ergodicity 
in the spin glass phase here, we have also studied the variation of the annealing time $\tau$ 
in the finite temperature Suzuki-Trotter Hamiltonian dynamics for $T(t) = T_0(1 - \frac{t}{\tau})$ and ${\Gamma}(t) = {\Gamma}_0(1 - \frac{t}{\tau})$ 
to reach a desired low value of the free-energy (corresponding to a very low, but finite, 
values of $T$ and $\Gamma$ to avoid singularities in the effective interaction flipping 
dynamics). Here $T_0$ and ${\Gamma}_0$ values of course correspond to the para phase. 
For such annealing through the ergodic region we have found $\tau$  to be fairly
independent of the system size $N$. However, it clearly starts growing with $N$ as one enters the  
nonergodic region (see~Fig.~\ref{phase_diagram}b). 

We believe, the numerical results reported here for the quantum SK model 
establishes the nature of the earlier conjectured~\cite{sudip-ray} ergodicity in 
the model and its role in quantum annealing~\cite{sudip-bikas,sudip-ttc-book,sudip_atanu_14} of
the SK model. It is also possible that the crossover region shrinks as $N \to \infty$. Indeed
there are several publications~\cite{sudip_Goldschmidt,sudip_sachdev,sudip_young_17} which contradict our conjecture
and suggest these results to be due to the finite size effects in the numerical simulations (of course, the paper by Read \textit{et al.} 
suggests ergodicity or absence of replica symmetry breaking as $T \to 0$). 
The same criticisms are also applied in~\cite{sudip_young_17} to the experimental and numerical observations~\cite{sudip-yao} for 
``scrambling'' or ergodicity in this system at low enough temperatures. Even if these effects are due to finite 
system size and the ergodic region becomes narrower with increasing system size, it is important to study such ``finite size scaling`` like   
behavior of the annealing dynamics, so that one can perhaps extrapolate properly the finite size annealing 
results as the system size approaches the macroscopic limit. Such an extrapolation scheme, if formulated properly, will be 
extremely useful for the quantum annealing machines (like D-wave~\cite{sudip_Denchev}) already developed and for developing the
quantum machine learning algorithms~\cite{sudip_Zecchina}.


\acknowledgements
We are grateful to Arnab Chatterjee, Arnab Das, Sabyasachi Nag, Purusattam Ray and Parongama Sen for their comments and suggestions. Bikas 
Chakrabarti gratefully acknowledges his J. C. Bose Fellowship (DST) Grant. Atanu Rajak acknowledges
financial support from the Israeli Science Foundation Grant
No. 1542/14.


\end{document}